\documentclass[aps,pra,preprint,groupedaddress,showpacs]{revtex4}

\usepackage{amsmath}
\usepackage{graphicx}

\begin{document}

\title{Semi-classical Over Barrier Model for low-velocity ion-atom 
charge exchange processes}

\author{Fabio Sattin }

\email{sattin@igi.pd.cnr.it}

\affiliation{Consorzio RFX, Associazione Euratom-ENEA , Corso Stati Uniti 4, 
35127 Padova, Italy}

\begin{abstract}
We develop an Over Barrier Model for computing charge 
exchange between ions and one-active-electron atoms at low impact 
energies. The main feature of the model is the treatment of the 
barrier crossing process by the electron within a simplified quantum 
mechanical formulation which takes into account: (a) the probability 
of electron reflection even for over-barrier scattering, and (b) the 
discreteness of the receiving atom's quantum levels which strongly 
suppress captures far from the resonance condition. It is shown that inclusion of these effects
yields a fairly good prediction of experimental data. We  discuss also 
the probability of electron re-capture by the target.
\end{abstract}
 
\pacs{34.70+e, 34.10.+x}

\date{\today}

\maketitle

\section{Introduction}
Charge exchange processes between slow atomic particles are of great 
importance in plasma physics and astrophysics. By ``slow'' 
we mean that the interparticle velocity is smaller than the classical velocity 
of the exchanged electron. \\ 
While only quantum mechanical (QM) methods can give really accurate 
computations of all of the basic quantities for these processes, i.e. 
total and partial or differential cross 
sections, less precise but simpler methods can still be highly valuable
when only moderate accuracy is sought. In the medium-to-high impact velocity 
range most preferences go to the Classical Trajectory Monte 
carlo (CTMC) method, which is also more and more often successfully 
applied also to the low velocity range (see e.g. \cite{rakovic,schultz} for a 
discussion and some recent improvements on this subject). However, the 
CTMC method has two disadvantages: (i) it is entirely numerical in character, 
thus somewhat masking the underlying physics; (ii) it relies on large 
numbers of simulations, thus being rather time-consuming. For these 
reasons, analytical or semi-analytical methods can still be useful. 
Over barrier models (OBM) are an example of these models. They are known 
since a long time \cite{ryufuku} and are still being improved to 
include as much physics as possible 
\cite{niehaus,ostrovsky,jpb,pra00,pra01}. \\
In this work we present a new version of the OBM. It is based upon the papers 
\cite{ostrovsky,pra00,pra01}, but is more than a simple refinement: we 
adopt here the approach of \cite{jpb}, assuming that it is always possible to improve 
any classical model by turning to a mixed description where some terms 
are computed using quantum mechanics. The main defect of model \cite{ostrovsky} 
is that it very often predicts too large a capture probability. To 
cure it, we were forced in works \cite{pra00,pra01} to artificially reduce 
the capture probability by arbitrarily reducing the capture region. We 
show here that similar results can be achieved in a more 
self-consistent way if the potential--barrier--crossing process by the electron is 
described as a quantum mechanical process. In particular, two typical 
QM features are taken into account: (i) the fraction of electrons crossing the barrier 
from the target atom to the projectile, which classically is a term of
exclusively geometrical origin, must be corrected by a factor $f_{t} < 
1$, accounting for the fact that a flux of quantal objects 
impinging on a potential hill suffers partial reflection even if 
their kinetic energy is larger than the hill's height.
(ii) Furthermore, within the classical picture the flux of electrons to the projectile is a 
continuos stream while, quantum--mechanically, it is a resonant process, 
occurring only when the conditions are satisfied for which the electron 
binds to a quantized energy level. We try to implement 
this feature by adding a modulation term, $w$, to the capture 
probability, near zero far 
from the resonance condition. \\
It is important to notice that, although QM 
corrections are empirically added to find convergence with 
experiments and/or other computations, no fitting parameters 
are added: instead, any new parameter needed is estimated on the basis of 
(semi)quantitative reasoning: once we accept the classical--quantal 
mixed description, the model is entirely self-consistent.  
Another important point to stress is that our goal 
is to merge the QM treatment within the classical one without burdening 
too much the resulting computations. We shall show that through a drastical simplification of the
QM computations we are able to have at the same time rather 
accurate results written in terms of simple formulas.  \\
Finally,  we try to implement in a consistent fashion within the 
model the possibility by the target of re-capturing the electron once it 
has been bound to the projectile. The importance of this effect on 
the effective capture efficiency, and the limitations of the approach 
adopted, will be briefly discussed in the appendix. 
 
\section{Description of the OBM}
\begin{figure}
\includegraphics[scale = 0.6]{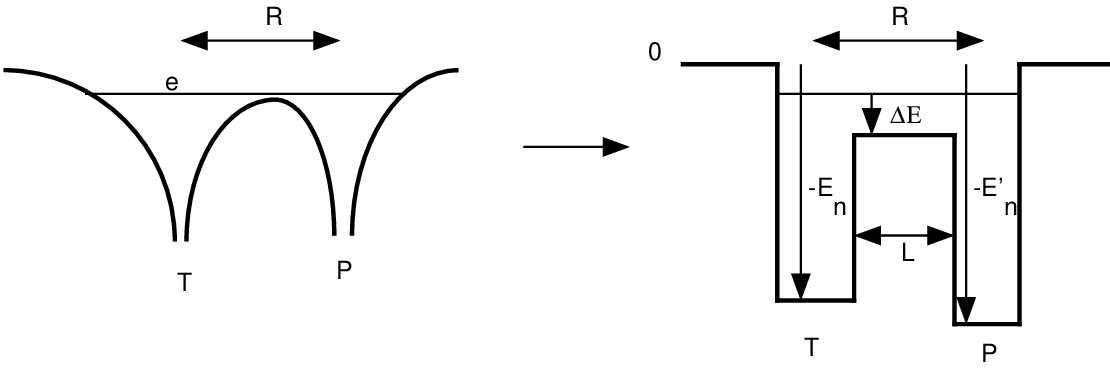}
\caption{\label{fig:uno} Pictorial view of the potential along the internuclear axis 
as it is (left figure) and how it is approximated by the rectangular 
square approximation (right figure).}
\end{figure}

\begin{figure}
\includegraphics[scale = 0.6]{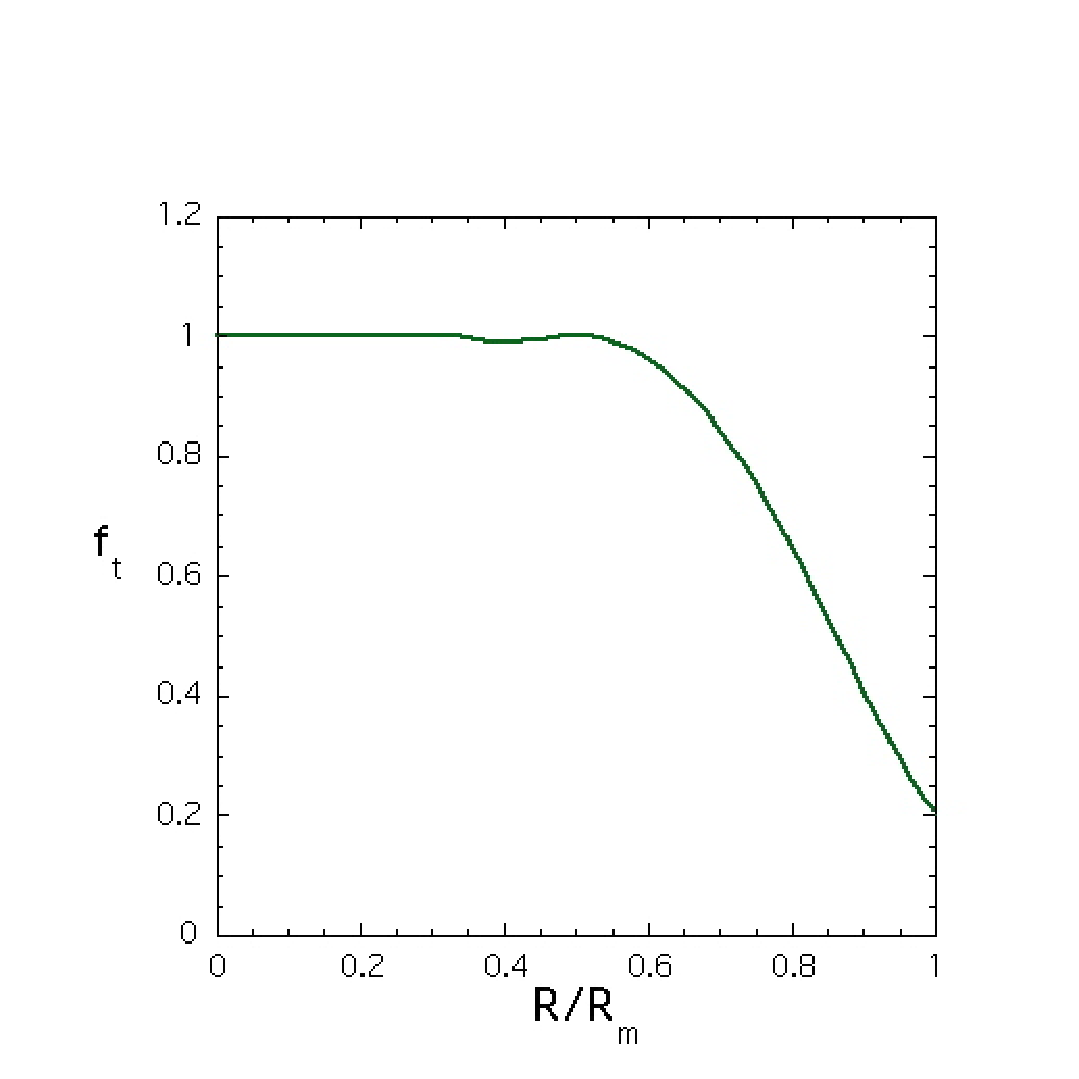}
\caption{\label{fig:due} Transmission factor $f_{t}$ (Eq. \ref{eq:trasmissione})
for H - H${}^{+}$ collisions. Notice that the probability of 
transmission falls well below one already for values of $R$ rather far from 
the maximum allowed capture distance $R_{m}$ (Eq. \ref{eq:rmassimo}).} 
\end{figure}

\begin{figure}
	\includegraphics[scale=0.6]{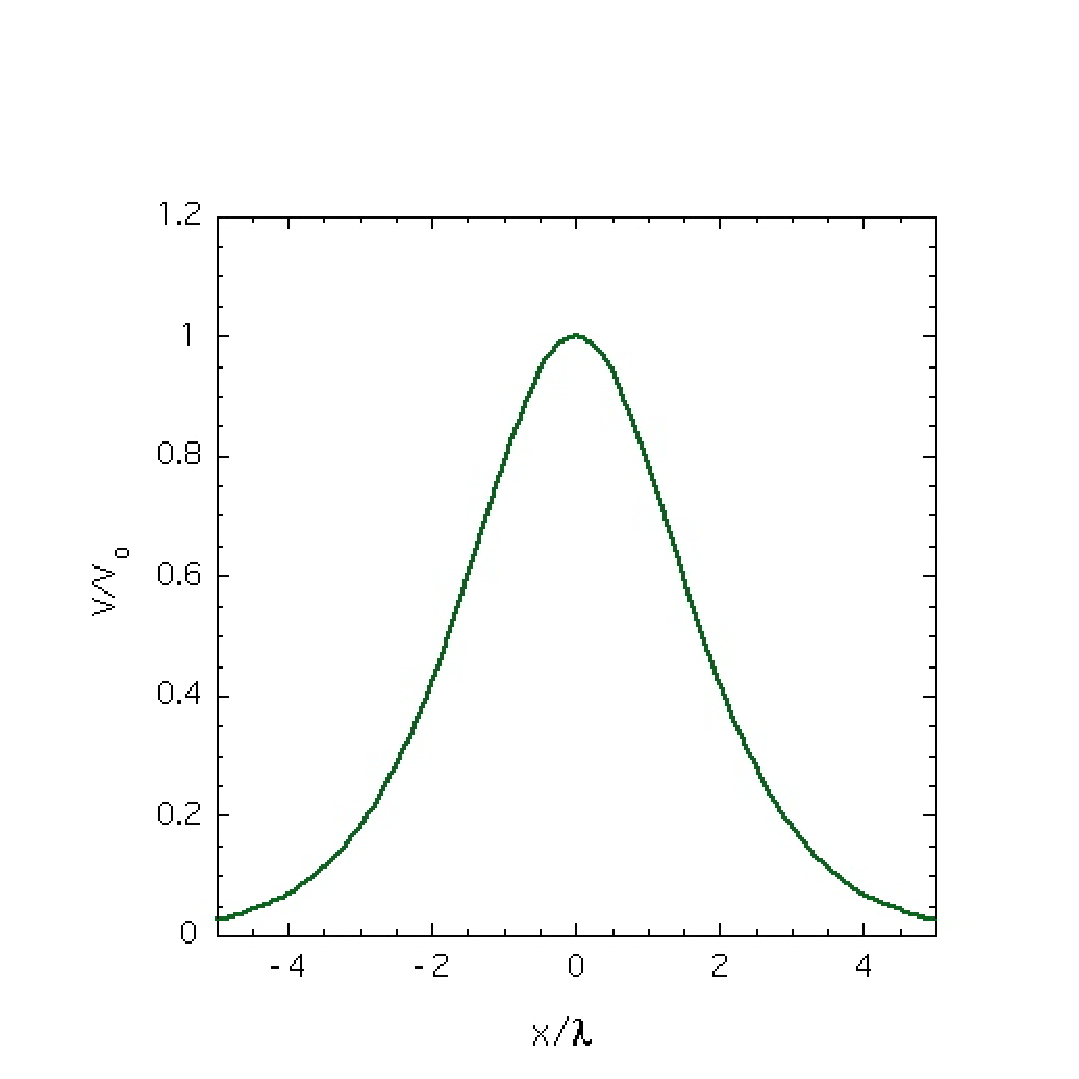}
	\caption{\label{fig:bell} Eckart potential (Eq. \ref{eq:eckart}).}
\end{figure}

We consider a scattering experiment between a nucleus {\bf T} with an 
active electron {\bf e}, and a projectile nucleus {\bf P}. Let {\bf r} 
be the electron position relative to {\bf T} and {\bf R} the relative 
distance between {\bf T} and {\bf P}. Let us further consider all the 
three particles lying within the same plane ${\cal P}$. We label the 
direction along the internuclear axis as the {\it z} axis and describe 
the position of the electron using cylindrical coordinates $(\rho, z, 
\phi \equiv 0)$. The two nuclei are considered as approaching at a 
velocity very small if compared to the orbital electron velocity. 
The total energy of the electron is
\begin{equation}
\label{eq:energiatotale}
E(R) = {p^{2} \over 2 } + U(z,\rho,R) = {p^{2} \over 2 }
-{ Z_{t} \over \sqrt{\rho^{2}+z^{2} } }- {Z_{p} \over \sqrt{\rho^{2}+(R-z)^{2}}} 
\quad .
\end{equation}
where $Z_{p}$ and $Z_{t}$ are the effective charge of the projectile and  
of the target seen by the electron, respectively (we are 
considering hydrogenlike approximations for both the target and the 
projectile). Atomic units are used unless otherwise stated. Notice 
that our reference frame is non inertial, so other terms as Coriolis 
and centrifugal force should arise. We discard them on the basis of 
the low-velocity approximation.\\
As long as the electron is bound to {\bf T} we can approximate $E$ by
\begin{equation}
\label{eq:energiarelativa}
E(R) \doteq - E_{n} - {Z_{p} \over R}  
\end{equation}
with $E_{n} > 0$ the unperturbed binding energy of the electron to {\bf 
T}.  \\
On the plane ${\cal P}$ we can draw a section of the equipotential 
surface 
\begin{equation}
\label{eq:equipotenziale}
 U(z,\rho,R) = E(R) = - E_n  - {Z_p \over R} \quad . 
\end{equation}
The set of points $(\rho, z)$ that satisfy this equation mark the limits of the region 
classically allowed to the electron. When $R \to \infty$ this region 
is disconnected into two circles centered around each of the two nuclei. 
As $R$ diminishes the two regions can eventually merge. It is the opening of the 
equipotential curve between {\bf T} and {\bf P} which leads to a 
leakage of electrons from one nucleus to another, and therefore to 
charge exchange. It is possible to solve Eq. (\ref{eq:equipotenziale}) 
in the limit of vanishing width of the opening $(\rho \equiv 0)$, and 
find:
\begin{equation}
\label{eq:rmassimo}
R_{m} = { (\sqrt{Z_{t}} + \sqrt{Z_{p}})^{2} - Z_{p} \over E_{n} } \quad .
\end{equation} 
In the region of the opening the potential has a saddle structure: 
along the internuclear axis it has a maximum at
\begin{equation}
\label{eq:saddle}
z = z_{0} = R { \sqrt{Z_{t}} \over \sqrt{Z_{p}} + \sqrt{Z_{t}} } 
\end{equation} 
In our work, following \cite{ostrovsky}, we assume that the electron 
is in a low angular momentum state (e.g. an $s$ states), thus unperturbed classical electron 
trajectories can be visualized as straight segments in the radial 
direction, starting from the target nucleus. We assume also that the qualitative shape 
of the trajectory is not changed by the collision: the electron free 
falls towards {\bf P}. \\
Charge loss occurs provided that the electron is able to cross the 
potential barrier. Let $W$ be the probability for the electron to be 
still bound to {\bf T} at time $t$. We write its rate of change as
\begin{equation}
\label{eq:rate}
{dW(t) \over dt} = - N_{t}  {1 \over T_{t}} f_{t} W(t) 
\end{equation}
Notice that we have defined the 
probability for an electron to be bound at {\bf P} as $ 1 - W(t)$, 
thus ruling out the possibility of ionization, which is, however, 
small for low-energy collisions.\\
Let us now explain the meaning of the terms in the right hand side of 
this equation: we identify $N_{t}$ as the fraction 
of classical particles which at time $t$ leave the target as a 
consequence of their motion. It is simply a 
geometrical factor: assuming a uniform distribution of electron 
trajectories, it is equal to the ratio between the solid angle 
intercepting the opening and the total $4 \pi$ solid angle. The 
azimuthal integration is straightforward and thus
\begin{equation}
\label{eq:nomega1}
N_{t} = {1 \over 2} \left(1 - {z_0 \over \sqrt{z_{0}^2 + 
\rho_{m}^{2}}} \right)
\end{equation}
with $\rho_{m}$ half-length of the opening in the radial direction, 
root of
\begin{equation}
\label{eq:rhom}
E_n + {Z_p \over R} = {Z_t \over \sqrt{z_0^2 + 
\rho_m^2}} + {Z_p \over \sqrt{(R - z_0)^2 + 
\rho_m^2}}
\end{equation}
A useful approximation is to expand Eq. (\ref{eq:rhom}) and then 
Eq. (\ref{eq:nomega1}) in powers of $\rho_{m}/R$ and retain only terms 
up to the second order. This was justified in the original paper 
\cite{ostrovsky} by the need of accurately modelling far collisions. 
However, this approximation turned out to be a rather accurate one 
since close encounters are weighted in the final cross section by the 
(small) impact parameter, thus even a rough estimate is not so much
important.\\
With this approximation, the calculation is straightforward and we 
find
\begin{equation}
\label{eq:nomega2}
N_{t} = {1 \over 2} {\sqrt{Z_{p}/Z_{t}} \over \left( \sqrt{Z_{p}} + 
\sqrt{Z_{t}} \right)^{2} } \left[ \left( \sqrt{Z_{p}} + 
\sqrt{Z_{t}} \right)^{2} - Z_{p} - E_{n} R \right] \quad .
\end{equation}
The parameter $T_{t}$ is the classical period of the electron bound 
to {\bf T}. It accounts for the fact that if the classical phases 
within the ensemble of the electrons are randomly distributed, during 
the time interval $dt$ only the fraction $dt/T_{t}$ come through the 
opening. The period can be calculated using the semiclassical relation 
$T_{t} = 2 \pi n_{\text{ eff}}^{3}$ (see e.g. \cite{pra00}) with $n_{\text{eff}}$ an effective 
quantum number for the electron bound to {\bf T}, which can be 
computed from eq. (\ref{eq:energiarelativa}) through $E_{n} = 
Z_{t}^{2}/(2 n_{\text{eff}}^{2})$. The result is 
\begin{equation}
\label{eq:periodo}
T_{t} = 2 \pi \left({ Z_{t}^2 \over 2 E_{n}}\right)^{3/2} { 1 \over 
\left( 1 + {Z_{p} \over E_{n} R} \right)^{3/2} } \quad .
\end{equation}
The factor $ T_{t}^{0} = 2 \pi (Z_{t}^2/(2 E_{n}))^{3/2}$ is the 
unperturbed period. It is clear from eq. (\ref{eq:periodo}) that 
$T_{t}/T_{t}^{0} < 1$, and it was shown in \cite{pra01} that the 
average (or effective) value of this 
ratio is typically of the order of 0.5. \\
Finally, $f_{t}$ accounts for quantum mechanical corrections to the 
barrier crossing probability. Classically $f_{t} \equiv 1$, while a 
flux of quantum mechanical particles impinging on a potential barrier is 
reduced by a factor $f_{t} < 1$ even though the kinetic energy of 
the particles is larger than the height of the hill. 
In order to reduce computation of $f_{t}$ to an easily manageable form we must replace the true 
potential profile with a model barrier whose transmission factor can 
be analytically computed. The choice of the model potential can be 
crucial for final results, so we have tested two candidates. 
The former choice is the rectangular barrier;
in fig. (\ref{fig:uno}) we draw a schematic picture of the potential profile 
along the internuclear axis: as it is (left-hand figure), and as it is 
approximated (right-hand figure). 
The horizontal line labelled by ``e'' marks the energy of the electron.  
The zero of the potential  well associated at the target is chosen so 
that the binding energy of the electron is equal to its unpertubed 
value $E_{n}$, and analogously $E'_{n'}$ is the binding 
energy to the projectile (till now undefined). The potential 
barrier between the two nuclei is depressed so that the electron is 
able to cross it with a kinetic energy $\Delta E$. The width of the 
potential barrier is set to $L$, yet undefined but of the order of 
the internuclear distance $R$. The transmission factor TF for an electron 
coming from the left of the potential hill is 
\begin{equation}
\label{eq:trasmissione}
\text{TF} = \frac{e^{-\frac{i }{2}\,L\,\left( k - 2\,q + s \right) }\,4\,k\,q}
  {
     k\,\left[ \left( q + s \right)  + e^{2\,i \,L\,q}\, \left( q - s 
     \right) \right] +  
      q\,\left[ \left( q + s \right)  - e^{2\,i \,L\,q}\, \left(q - s 
      \right) \right]   
   }
\end{equation}   
with $k = \sqrt{2 E_{n}} ,\, q = \sqrt{2 \Delta E} \, , \, s = \sqrt{2 
E'_{n'}}$ electron momenta respectively in the {\bf 
T} potential well, in the potential barrier, and in the {\bf P} 
potential barrier. Of course, the relation holds 
\begin{equation}
\label{eq:ft}
f_{t} = |\text{TF}|^{2}  \quad .
\end{equation}
The binding energy to the projectile can be calculated by considering 
that, when {\bf e} is bound to {\bf P}, its energy is $E(R) = 
-E'_{n'} - Z_{t}/R$, with $ E'_{n'} = Z_{p}^{2}/(n')^{2}$. At the capture radius this expression and that 
given by eq. (\ref{eq:energiarelativa}) must be equal, thus
\begin{equation}
\label{eq:eprimon}
E'_{n'} = E_{n} + { Z_{p} - Z_{t} \over R} \quad .
\end{equation}
We compute $q$ along the internuclear axis:  by using eqns. 
(\ref{eq:energiatotale},\ref{eq:energiarelativa},\ref{eq:saddle}), it 
is quite easy to work out
\begin{equation}
\label{eq:qu}
q = k \left( {R_{m} \over R} - 1 \right)^{1/2} \quad .
\end{equation}
It remains to estimate $L$. Of course, given the way eq. 
(\ref{eq:trasmissione}) has been derived, only a semi-quantitative estimate 
is required. By a straightforward application of the virial theorem, 
one finds 
\begin{equation}
\langle { p^{2} \over 2 } \rangle  \doteq {1 \over 2} Z_{t} \langle { 1 \over 
S_{t}  } \rangle  \approx {1 \over 2} { Z_{t}  \over \langle S_{t} \rangle }
\end{equation}
and
\begin{equation}
\langle { p^{2} \over 2 } \rangle  -   Z_{t} \langle { 1  \over S_{t} } \rangle 
= - E_{n}
\end{equation}
with $ 2 \, \langle S_{t} \rangle $ average width of the potential well. A similar relation 
holds for the potential well centered around the projectile. 
From the two previous equations we find 
$ \langle S_{t} \rangle = 1/2 (Z_{t}/E_{n})$, $ \langle S_{p} \rangle = 
1/2 (Z_{p}/E'_{n})$, thus,  
\begin{equation}
\label{eq:elle}
 L = R -  ( <S_{t}> + <S_{p}> ) = R - {Z_{t} \over 2 E_{n}} - { Z_{p} \over 2 E'_{n}} \quad .
\end{equation}
Of course, we set $L = 0$ when the right-hand side of the 
equation above is lesser than zero. By taking a glance at eqns. 
(\ref{eq:trasmissione},\ref{eq:elle}), one can guess that  
the effective number of captures 
$f_{T} N_{T}$ is strongly suppressed already 
for $R < R_{m}$: this is exactly what found in CTMC 
simulations (see \cite{pra00}). 
In order to take an insight at what the transmission factor looks 
like, we plot in fig. (\ref{fig:due}) $f_{t}$ given by Eq. 
(\ref{eq:trasmissione}) for  H - H${}^{+}$ 
scattering. 
In the same way as the laws of quantum 
mechanics prevent a fraction of electron to be captured even when it 
would be classically allowed, they also--through tunnelling--would make it 
possible for some electrons to be captured even at internuclear 
distances $ R > R_{m}$. However, it is easy to show that this 
correction to the total capture probability is very small, and thus 
we will neglect it. \\
This model potential is extremely useful for calculations but one 
could wonder if it is too drastic an approximation, particularly in 
view of the fact that it is not smooth. 
As a test, we have replaced the rectangular barrier with an 
Eckart  potential \cite{eckart}
\begin{equation}
\label{eq:eckart}
V(x) = 4 V_{0} {  e^{ x  \over \lambda } \over (1 + e^{ x  \over \lambda})^{2} } \quad ,
\end{equation}
which is instead a smooth, bell-shaped curve (see fig \ref{fig:bell}).
 The transmission factor for this 
potential is analytically computable and, for a particle of momentum $p$:
\begin{equation}
	\label{eq:feckart}
f_{t} = { \cosh(4 \pi p \lambda) - 1 \over 
           \cosh( 4 \pi p \lambda) + \cosh( 4 \pi \sqrt{ 2 \lambda^2 V_{0} - 
           1/16})} \quad .
\end{equation}
The parameters $\lambda, V_{0}, p$ can be straightforwardly related to the parameters $L, 
k, q, s$ given above. The quantity $f_{t}$ in (\ref{eq:feckart}) is a mildly 
varying function of $\lambda$ but is, unfortunately, a strong 
function of $E/V_{0}$, going steeply from zero to one as this ratio 
crosses the unity. This is a deprecable feature since, as it appears 
clear from the equations (\ref{eq:eprimon}-\ref{eq:elle}) above, we 
can give only semiquantitative estimates-therefore likely to have 
a large variability-for all the quantities involved.\\  

A main difference between the classical picture 
and the quantum-mechanical one is that the former depicts the capture 
process as a continuous flow. On the contrary, quantization rules 
forbid the electronic flow from one nucleus to the other unless some 
resonance conditions are satisfied: by using the relation 
(\ref{eq:eprimon}),  we obtain
\begin{equation}
\label{eq:rvsn}
R(n') = { Z_{p} - Z_{t} \over { {1 \over 2} {Z_{p}^{2} \over 
(n')^{2}} - E_{n} } } \quad ,
\end{equation}
that is, captures should be allowed only close to these values of 
$R$, in correspondence of integer values $n'$. \\
In this work we want to implement an algorithm to take into account 
the damping of capture probability far from these resonances, 
persuaded that this should drastically improve the predictions of the 
model. \\
We choose to implement phenomenologically this feature by modulating 
$f_{t}$ with a weight function $w$ centered around the values 
$R(n')$. For $w$ we choose a sum of Gaussians
\begin{equation}
\label{eq:peso}
w(R) = \sum_{n'} \exp\left[- \left({ R - R(n') \over \Delta R(n')/2} 
\right)^{2}\right] \quad .
\end{equation}   
It is necessary to find a reasonable value for $\Delta R(n')$. For this we 
resort to the indeterminacy relations: infact, $ \Delta R \to 0$ means 
that the energetic levels for the projectile are sharply defined, 
while any finite value for $\Delta R$ means that they are defined 
only within an energy range $\Delta E$. We suppose that the usual 
indeterminancy relations hold: $\Delta E \times \Delta t \doteq 1/2$.
Within the straight--line trajectory approximation,
$ \Delta t \approx \Delta R/u$ while, using again Eq. 
(\ref{eq:eprimon}) and differentiating with respect to $R$, we get 
$ \Delta E \approx (Z_{p} - Z_{t})/R^{2} \Delta R$. Collecting the 
above expressions,
\begin{equation}
\label{eq:deltar}
\Delta R (n') \approx \sqrt{ {1 \over 2} { u R^{2}(n') \over Z_{p} - 
Z_{t}} } \quad .
\end{equation}
The above expression breaks down when $Z_{p} = Z_{t}$. In that case we will 
not adopt this approach, instead we wil assume uniform probability of 
capture all along the internuclear distance: $ w \equiv 1$. 
Notice that, with the above definitions, it seems that we get as a bonus
also partial probabilities for capture into well defined 
quantum numbers: if we are interested in captures into state $m$,
it is enough to truncate the sum $w$ (Eq. \ref{eq:peso}) to the 
single term corresponding to $n' = m$. As we shall see later, this 
picture however does not hold. \\
Eq. (\ref{eq:rate}) can be formally integrated till the end of 
collision: 
\begin{equation}
\label{eq:w}
W(t = \infty) = \,\exp \left[ - \int_{-\infty}^{\infty} \left( { w f_{t } 
\, N_{t} \over T_{t}} \right) \, dt  \right] \quad .
\end{equation}
The capture probability is $ P = 1 - W(\infty)$. 
We assume a straight-line trajectory for the projectile: $ R = \sqrt{b^{2} + (u 
t)^{2}}$, with $b$ impact parameter and $u$ its velocity. Total charge exchange cross 
section is thus 
\begin{equation}
\sigma = \, 2 \pi \int b \,P \, db   \quad .
\end{equation} 
Equation (\ref{eq:w}) cannot be analitically integrated unless we 
make further simplifications. However, it is quite easily numerically 
integrated by any standard mathematical software package.

\section{Results}
\begin{figure}
\includegraphics[scale = 0.6]{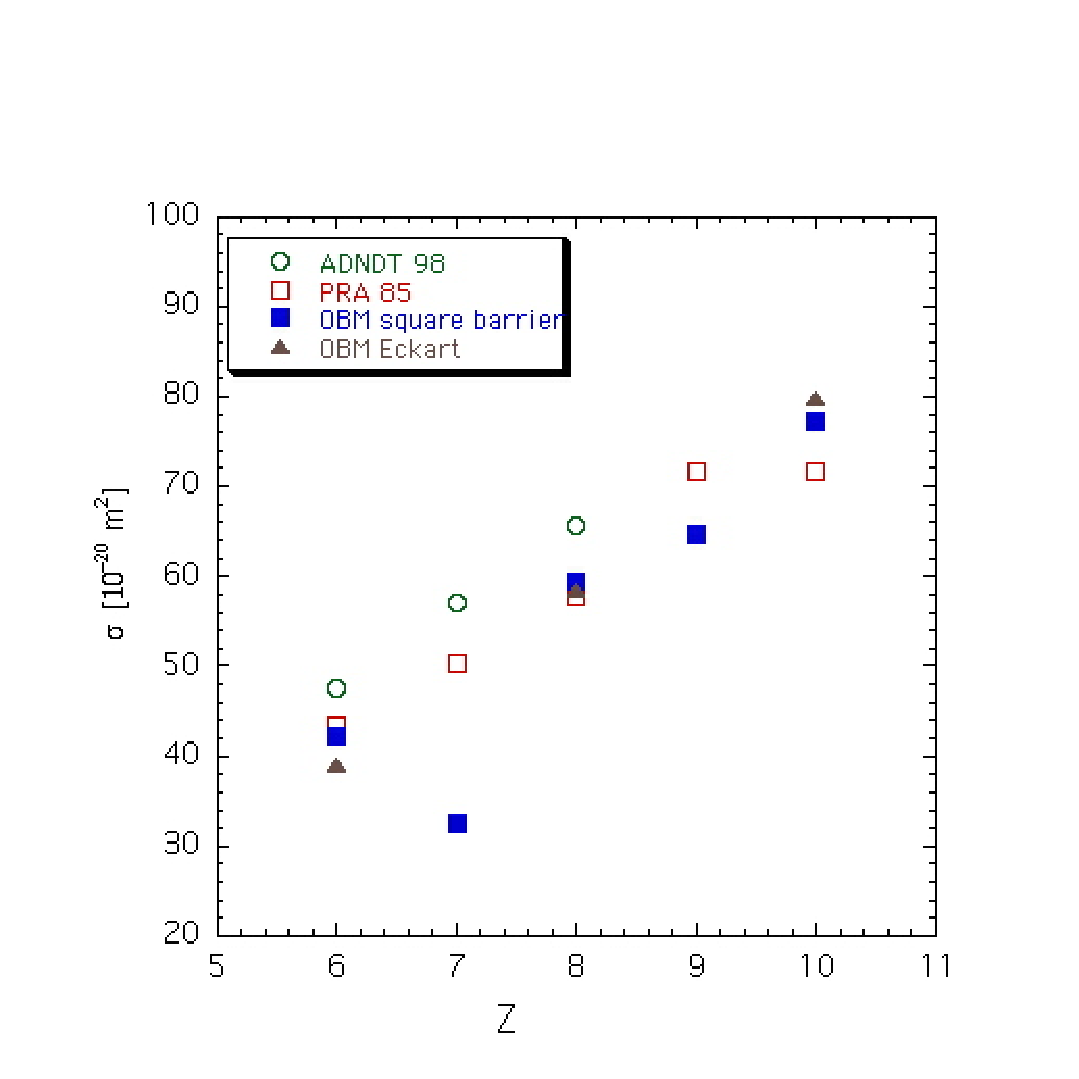}
\caption{\label{fig:tre} Charge exchange cross section versus projectile charge.
Open circles, data from ref. \cite{harel}; 
open squares, data from ref. \cite{meyer}; full squares, OBM results using rectangular barrier potential; 
triangles, OBM results using Eckart potential (Eq. \ref{eq:eckart}). }
\end{figure}

\begin{figure}
\includegraphics[scale = 0.6]{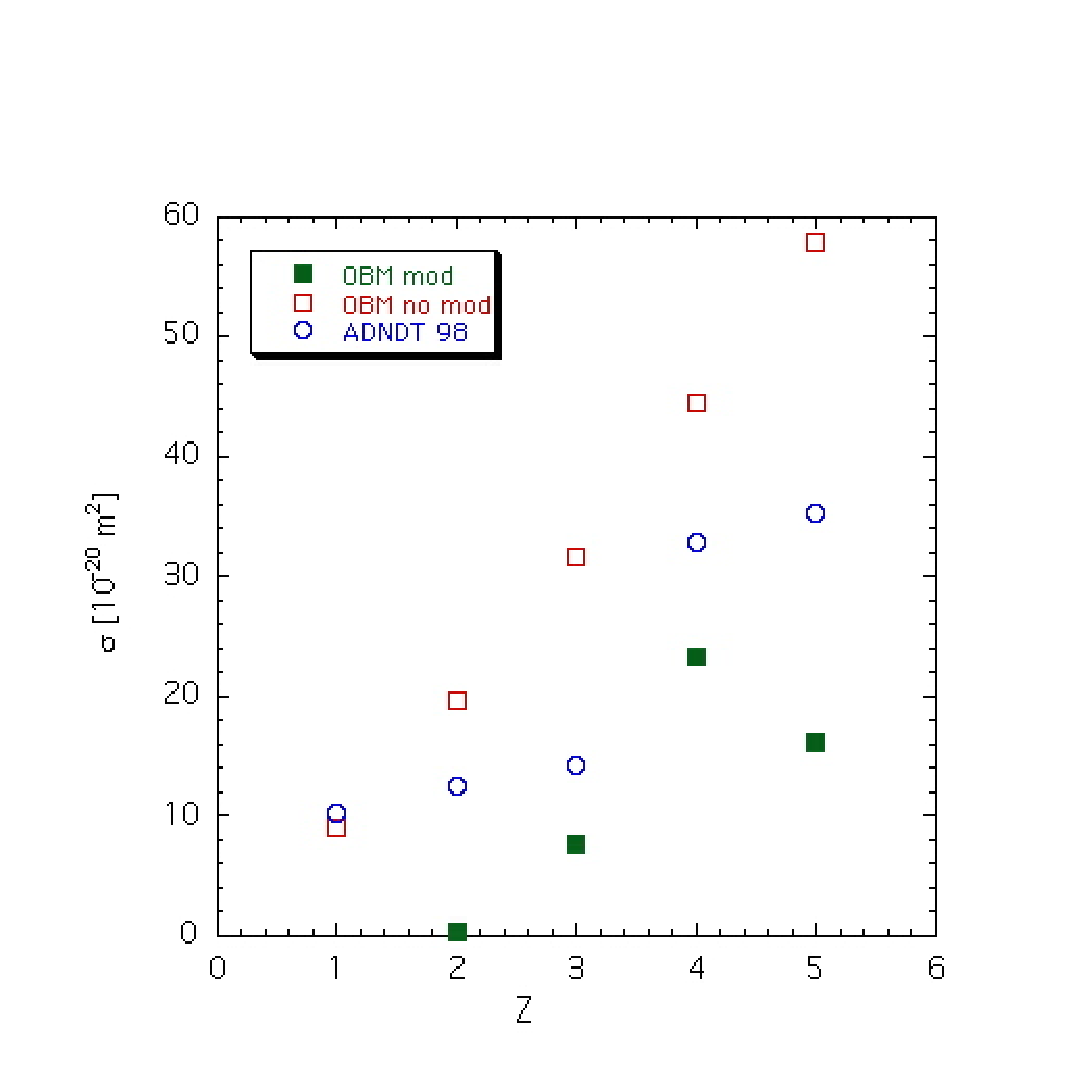}
\caption{\label{fig:quattro} Charge exchange cross section versus projectile 
charge,$ 1 \leq Z \leq 5$.
Open circles, data from ref. \cite{harel}; 
full squares, OBM results using rectangular barrier potential; 
open squares, same as full squares but without modulation of the 
transmission factor.}
\end{figure}

\begin{figure}
\includegraphics[scale = 0.6]{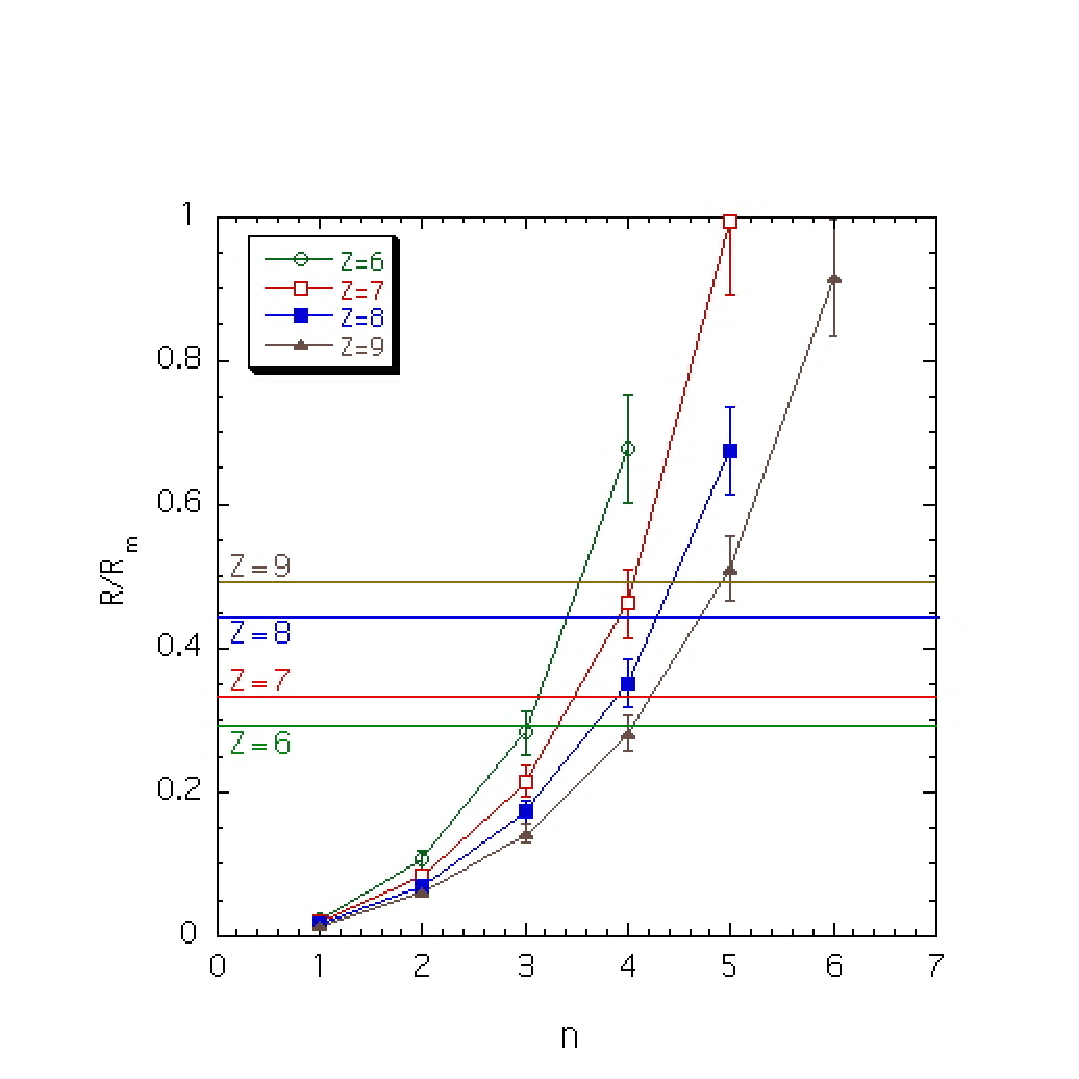}
\caption{\label{fig:cinque} Normalized capture distance {\it versus} 
quantum number, for some highly-charged ions. Horizontal lines mark, 
for each ion the distance at which the attenuation factor (\ref{eq:ft}) is 
reduced to one half. }
\end{figure}

We will benchmark the model against the experimental results from ref. 
\cite{meyer} and the theoretical ones coming from the molecular 
approach simulation of ref. \cite{harel}.  In fig. (\ref{fig:tre}) we 
show the results for impact between 
multicharged hydrogenlike ions and  ground state hydrogen. 
In all case impact velocity is about 1/2 (it is rigorously so for numerical results, while 
in experiments $ 0.49 \leq u \leq 0.51 $). 
Meyer et al.  \cite{meyer} published also results from other non-hydrogenlike ions, 
with the same charge states; since, for a given charge, cross sections are not qualitatively different, 
we have limited to the present selection and got rid of the complications involved 
with the definition of effective charges. \\
Let us now discuss the results of fig. (\ref{fig:tre}).  
The OBM with the choice of the rectangular barrier as model potential 
(black squares) fairly well overlaps experimental data for all cases but 
N${}^{7+}$. The reason lies in a combination of effects, which we 
will discuss below.  Let us comment instead, the results from the 
other simulation (black triangles): they have been computed using 
Eckart potential  (Eq. \ref{eq:eckart}). Even if at first 
sight these results look pretty accurate, things are not that good. 
Actually, in order to reach this accuracy, we have had to increase 
the barrier height $V_{0}$ so as to allow only grazing incidence, $ 
E = V_{0}$ (or $q \approx 0$, using notations of rectangular barrier 
potential). This, since Eckart potential has a strong transmission 
factor for $E/V_{0}$ even slightly larger than one. Using the same 
conditions as for the rectangular barrier would yield an overestimate 
of the true cross section. \\
Let us now consider fig. \ref{fig:quattro}: it is alike fig. \ref{fig:tre} 
but with $ 1 \leq Z \leq 5$. Only data from ref. \cite{harel} are 
available for these ions.  In this case we have computes $\sigma$ 
using only $f_t$ from Eq. (\ref{eq:ft}) but with $w$ given by Eq. 
(\ref{eq:peso}) (solid circles) or with $w \equiv 1$ (empty circles). 
The only exception being hydrogen, for which only $ w = 1$ can be used.
It is apparent that some sort of modulation of the transmission factor is needed 
in order to avoid an overestimate of $\sigma$. The effect of the 
modulation is to underestimate the cross section, but some charge states 
are more damped than others. It is the same effect, but enhanced, 
seen in fig. \ref{fig:tre}.  The reason of the failure of the model 
in these cases lies in a balance between attenuation factors and 
resonances positions. In order to better visualize it, let us 
consider fig. \ref{fig:cinque}: there, we have plotted, 
for ions with charge $ 6 \leq Z \leq 9$, the position of each 
resonance $R(n)$ [normalized to the maximum capture distance $R_{m}$ 
given by Eq. (\ref{eq:rmassimo})] versus the quantum number $n$. The bars 
superposed to the points mark the width of the resonance: they are 
the $\Delta R(n)$'s given by Eq. (\ref{eq:deltar}). The horizontal 
lines label the distance at which the capture probability is reduced 
to 1/2 because of the factor (\ref{eq:ft}). Thus, capture is effective 
only into quantum numbers whose resonance position lies below this line.       
We compare, for example, carbon (Z = 6) to nitrogen (Z = 7): in both 
cases, attenuation $f_{t}$ limits capture effectively to the first three 
states, and impact parameter considerations lead to suggest that 
$n = 3$ is the dominant state. However, the resonance width of $n = 
3$ is lesser for nitrogen than for carbon, thus explaining the reduced 
capture cross section.  Similar considerations, but a lesser extent, 
hold for $ Z = 8$ and $ Z = 9$. It is clear that
slightly charged ions, for which captures only in few states are 
possible, are particularly sensitive to this effect.  Different is the case for 
hydrogen, where, as explained in the previous section, no 
considerations of resonances apply. Notice, however, that 
oscillations in capture cross sections with the charge, due to discreteness of quantum 
levels, are not an artifact of the model but, albeit with different 
features, have been experimentally observed \cite{ryufuku}. 

\section{Conclusions}
In this work we have presented a version of Over Barrier Model for 
computation of charge exchange cross sections.
The model presents some unusual features, such as the inclusion within 
its classical background of quantum elements. 
We showed it to be quite reliable in a well defined range of 
parameters, namely in the low-energy, medium-to-high-projectile charge 
region, where it presents noticeable improvements with respect to 
other similar algorithms.  \\
The reasons for poorer performances in other parameters space regions 
have been clearly identified and some insights about possible further 
improvements are possible: the first candidate to work on
to reach an overall fairly good accordance with experiment is the 
modulation factor $w$ (Eq. \ref{eq:peso});
presently, however, we are not able to guess if a general optimal expression is 
possible, valid for all projectile-target combinations. 
\appendix

\section{}
It is straightforward to adapt Eq. (\ref{eq:rate}) to accomodate the 
possibility for the captured electron to return to the target nucleus: 
it is enough to add a term
\begin{equation}
\label{eq:rate2}
{dW(t) \over dt} = - N_{t}  {1 \over T_{t}} f_{t} W(t) + N_{p} 
{1 \over T_{p}} f_{p} (1 - W(t))
\end{equation}
The second term in the r.h.s. represents the 
flux of electrons which have previously been captured by  {\bf P} and 
that now cross a second time the barrier in the opposite direction, 
thus being re-captured by {\bf T}.  The definition of the parameters
$N_{p}, T_{p}, f_{p}$ is the same as given in the previous sections, 
see Eqns . (\ref{eq:nomega1},\ref{eq:periodo},\ref{eq:ft}), 
with the trivial exchange of the projectile with the target.\\      
Eq. (\ref{eq:rate2}) can be solved exactly in terms of quadratures. 
Here, we will show that it is possible from this equation to recover the result for symmetrical scattering: 
infact, in this case, $f_{t} = f_{p}\, ,\, N_{t} = N_{p} $ and $T_{t} = T_{p}$ and, 
by setting $ f_{t} N_{t}/T_{t} = f_{p} N_{p}/T_{p} = \varphi$ for 
brevity, and $ \int_{\infty}^{t} \varphi(\tau) \, d\tau = \Phi(t)$, we get
\begin{equation}
\begin{split}
\label{eq:psimmetrica}
P =& \,1 - e^{- 2 \Phi(\infty) }
      \left[ 1 + \int_{-\infty}^{\infty} \varphi(\tau) \, e^{2 \Phi(\tau)} \, d\tau \right] \\
        =& \,1 - e^{ - 2 \Phi(\infty) }
        \left[ 1 + {1 \over 2} \, \int_{-\infty}^{\infty} {d \over d\tau} e^{2 \Phi(\tau)} \, d\tau \right] \\
        =& \,1 - e^{- 2 \Phi(\infty)} \left[1 + {1 \over 2} \left( e^{2 
        \Phi(\infty)} - 1 
        \right) \right] \\
        =& \,{1 \over 2} \left[ 1 - e^{-2 \Phi(\infty)} \right] 
        \quad ,
\end{split}
\end{equation}   
and we recover the prescription for equal-charge scattering. \\
Of course, we can also recover easily the opposite 
limit, in which $Z_{p} >> Z_{t}$; in this case we can simply neglect the 
return term since $f_{p} N_{p} /T_{p} << f_{t} N_{t}/T_{t}$. \\
We have verified that the probability of recapture is negligible for 
all multicharged projectiles. The only exception is the $Z_{p} = 
Z_{t}$ scattering. However, we found that the use of the symmetrized 
capture probability (\ref{eq:psimmetrica}) does not improve results 
in this case, and actually degrades the performances of the model. 
The reason is that we are overestimating the probability of recapture 
in Eq. (\ref{eq:rate2}): it holds rigorously in the limit of 
asymptotically slow nuclear motion, $ u \to 0 $; for finite $u$, 
retardation effects should be implemented to take into account that 
electrons released by the projectile at time $t$ were captured at an 
earlier time $t'$. However, in the region of velocities where retardation 
effects can be neglected, other essential features of the model, such as 
the straight trajectory approximation, break down.

\end{document}